
%
\magnification=1200
\vsize=7.5in
\hsize=5in
\pageno=0
\footline={\ifnum\pageno=0{}\else\hfil\number\pageno\hfil\fi}

\hfuzz=5pt
\baselineskip 12pt plus 2pt minus 2pt
\centerline{\bf DYNAMICAL PROPERTIES OF A}
\centerline{\bf HALDANE GAP ANTIFERROMAGNET}
\vskip 24pt
\centerline{O. Golinelli,
Th. Jolic\oe ur,\footnote{*}{C.N.R.S. Research Fellow} and
R. Lacaze$^*$}
\vskip 12pt
\centerline{\it Service de Physique Th\'eorique\footnote{**}
{\rm Laboratoire de la Direction des Sciences de la Mati\`ere
 du Commissariat \`a l'Energie Atomique}}
\centerline{\it C.E.  Saclay}
\centerline{\it F-91191 Gif-sur-Yvette CEDEX, France}
\vskip 1.0in
\centerline{\bf ABSTRACT}
\vskip 1.0in
We study the dynamic spin correlation function of a spin one
antiferromagnetic chain with easy-plane single-ion anisotropy.
We use exact diagonalization by the Lancz\H os method for chains of
lengths up to N=16 spins. We show that a single-mode approximation
is an excellent description of the dynamical properties.
A variational calculation allows us to clarify the nature of the
excitations. The existence of a two-particle continuum near zero wavevector
is clearly seen both in finite-size effects and in the dynamical structure
factor. The recent neutron scattering experiments on the
quasi-one-dimensional antiferromagnet NENP are fully explained by our results.
\vskip 1.0in
\centerline{Submitted to: {\it J. Phys. C: Condensed Matter}}
\vskip 0.2in
\noindent November 1992

\noindent PACS No: 75.10J, 74.50M, 75.50E \hfill SPhT/92-135
\vskip 1.0in
\vfill
\eject
\noindent{\bf I. INTRODUCTION}
\medskip
It was first argued by Haldane [1,2] that generic spin-{\it S} one-dimensional
Heisenberg antiferromagnets have an excitation gap for integer {\it S}. This
picture is quite different from the usual, higher-dimensional, picture of
antiferromagnets with massless Goldstone magnons. On the theoretical side,
there is now convincing evidence from numerical studies [3-10] of the {\it S}=1
Heisenberg chain that it has a nonzero gap in the thermodynamic limit. Exact
diagonalizations by the Lancz\H os method have been able to reach 18 spins
and Monte-Carlo simulations extend the range to 32 spins. All these data are
suggestive of a Haldane gap. This leads to a ground state spin correlation
length
which is  finite and about six lattice spacings. Such a picture is in striking
contrast with the {\it S}=$1/2$ solvable chain which is gapless and whose
ground state has algebraically decaying correlations.
 A spin 1 Heisenberg model with biquadratic
exchange has also been discovered with an exactly solvable ground state and a
nonzero gap [11], thus reinforcing the belief in the Haldane conjecture.

On the experimental side there are several candidates to exhibit this quantum
gap. The first experimental evidence came from neutron-scattering on
CsNiCl$_3$ [12-15]. In this compound there are chains of  spin one Ni$^{2+}$
ions with superexchange through Cl$^{-}$ anions. There is however moderately
small interchain couplings that complicate the picture: this is revealed in
recent experiments on the related compound RbNiCl$_3$ [16]. The best candidate
so far seems to be Ni(C$_2$H$_8$N$_2$)$_2$NO$_2$ClO$_4$ (NENP) [17-20],
as shown by inelastic neutron-scattering (INS) and magnetization measurements.
The ratio of the interchain to intrachain magnetic coupling is estimated to be
$J^\prime /J \approx 4\times 10^{-4}$. No transition to N\'eel order is found
down to 1.2 K which is consistent with the hypothesis of a {\it finite}
zero-temperature correlation length. Nickel ions have spin 1 and are described
by the following anisotropic Heisenberg Hamiltonian [21]:
$$
H = J\sum_i {\vec S_i}\cdot {\vec S_{i+1} }+D\sum_i (S_i^z )^2.
\eqno(1.1)
$$
Best fit of INS gap values leads to $J/k_B = 43.5 K$ and easy-plane
$D/J=0.18$ [10].
There is also evidence for a smaller in-plane anisotropy that can be described
by adding to H a term $E\sum_i [(S_i^x)^2 -(S_i^y)^2]$. This perturbation is
small and will be mostly ignored in the remainder of this paper. Its
qualitative
role will be discussed in section V.

The first INS measurements [17-19] have concentrated on the neighborhood of
Q=$\pi$ (where Q is the wavevector along the chain) and showed the existence
of two gaps: one for the in-plane (IP) magnetic fluctuations and a higher
one for out-of-plane (OP) fluctuations. This splitting is due to the presence
of
a sizable easy-plane single-ion anisotropy $D$. The dispersion of the magnetic
excitations was then studied in the range $Q/\pi = 0.9-1.0$. Recent experiments
have extended our knowledge throughout the whole Brillouin zone [22].
Detailed theoretical work is required to test the hypothesis that the simple
model Hamiltonian (1.1) is able to reproduce the dynamical properties of NENP.
Two approaches have been followed up to now: small cluster numerical studies
[3-10,23] and the use of effective field theories [24,25]. The previous
numerical
studies were limited to spectral calculations or have ignored
the presence of in-plane anisotropy.

In this paper we present the results of a study of the dynamical structure
factor $S(Q,\omega )$ of chains of length up to 16 spins by means
of an exact diagonalization method.
We also apply a variational technique proposed in ref.[26] to the anisotropic
chain. This physically motivated method reproduces extremely well the
numerical results.
The results presented are valid in the absence of an external magnetic
field and in the zero temperature limit, i.e. for temperatures well below the
gap.
In sect. II we present some general properties
of the spin-1 chain with easy-plane single-ion anisotropy. In sect. III we
explain the methods used to obtain dynamical quantities. Sect. IV contains
our results. They are extremely well reproduced in terms of a physically
appealing single-mode approximation. The results of our variational
calculation are presented there and we use them to understand the
nature of the elementary excitations near $Q=0$ and $Q=\pi$.
We show in sect. V that they are in very
good agreement with existing neutron data and we suggest an additional test of
the theory. Sect. VI contains our conclusions.
\bigskip
\bigskip
\noindent{\bf II. GENERAL PROPERTIES}
\medskip
We briefly review some known results about the
isotropic Heisenberg spin 1 antiferromagnetic (AF) chain:
$$
H_0 = J\sum_{i=1}^{N} {\vec S_i}\cdot {\vec S_{i+1} }.
\eqno(2.1)
$$
The vectors $\vec S_i$ are quantum spin operators satisfying the SU(2)
rotation algebra with length ${\vec S_i}^2 = 2$. They are located along a
one dimensional lattice of N sites with periodic boundary conditions.
The exchange integral J is positive in the AF case.
For any finite lattice the ground state is a singlet and the higher-lying
levels
have energy increasing with increasing spin as is known rigorously [27].
Above the singlet ground state one finds a triplet state presumably for all
values of N. In an AF quantum magnet, in the case of broken symmetry one
expects
that in the thermodynamic limit the triplet becomes degenerate with the ground
state as do other states with spin S=2,3,\dots , in order to form the
degenerate ground state of the infinite volume limit.
This can occur only for dimension greater than 2 (or equal).
 Haldane has argued
that the spin-1 chain is quantum disordered and that the singlet-triplet gap
remains nonzero in the thermodynamic limit. This should be true for any integer
value of the spin. This argument is based on a mapping of the infinite-spin,
semiclassical limit of the quantum chain onto a O(3) nonlinear $\sigma$ model
[1]. In the spin-1 case of interest, numerical studies clearly points towards a
gap close to $\approx 0.41 J$ [3-10,23]. The physics of the O(3) nonlinear
$\sigma$  model is that of a triplet of massive bosons with nontrivial
scattering
properties. This suggests a very simple effective field theory: a free theory
of
three massive bosons [25]. Another route starting from an integrable model
leads
to an effective field theory of three massive fermions [24].

In such approaches, one
has to adjust the gap values which are no longer deducible
from the microscopic model but one can obtain simple
and definite prediction on dynamical quantities. However both are approximate
theories and their respective domain of validity is difficult to assess.
On the contrary, finite chains calculations offer unbiased theoretical
predictions provided one is able to carefully control finite-size effects.
Concerning the gap values this can be achieved by the use of the so-called
Shanks transformation, suited to the removal of exponential transients in
sequences of finite-chain data [28]. It has been found that the convergence
towards infinite-volume is very good [6,10,23] as expected since we are dealing
with a massive theory.

As shown with the Perron-Frobenius theorem [29], the lowest-lying
triplet has wavevector $Q=\pi$, while the singlet ground state has $Q=0$.
Most theoretical and experimental studies have concentrated on this $Q=\pi$
part
of the spectrum. In the nonlinear $\sigma$ model, there are no bound states
and,
if we believe that it is the effective theory of the spin 1 chain, this implies
that the $Q=0$ gap is due to two massive $Q=\pi$ particles and thus twice the
$Q=\pi$ gap [2]. This has been checked by a quantum Monte-Carlo simulation
in the isotropic case [8] and Lancz\H os studies [23] have shown that this
property persists in the presence of anisotropy (as long as it is not too
large).  As a consequence, the states near $Q=0$ should be members of a
two-particle continuum contrary to the states near $Q=\pi$ that should appear
as
long-lived well-defined excitations.
INS measurements for $Q=0$ [30] reveal a vanishing structure factor in this
regime. The curve of the lowest excited triplet at wavevector $Q$ is thus
bell-shaped but asymmetrical with respect to $Q=\pi /2$.

Let us now discuss the influence of easy-plane single-ion anisotropy, i.e.
adding to $H_0$ a term of the following form:
$$
H = H_0 + D \sum_i (S^z_i )^2 .
\eqno(2.2)
$$
The full rotational symmetry of Hamiltonian (2.1) is broken to rotational
symmetry around the z axis.
Only the z component $S^z=\sum_i S^z_i$ of the total spin is conserved.
There is also a discrete symmetry $S^z\rightarrow -S^z$ that is preserved
in the Hamiltonian (2.2).
As a consequence of the anisotropy, the first excited triplet state is split
into a higher-energy singlet $S^z =0$ and a lower-lying doublet $S^z =\pm 1$.
These three states retain their wavevectors unchanged ($Q=\pi$) with
respect to the $D=0$ case (by continuity)
since the Hamiltonian (2.2) still
possesses translational symmetry. The Haldane gap is split into two components:
one gap $G^{(-)}$ between the ground state (with $S^z =0$) and the doublet
$S^z =\pm 1$ and one gap $G^{(+)}$ inside the $S^z =0$ subspace between the
first
two levels. $G^{(-)}$ decreases  while $G^{(+)}$ increases with increasing D as
shown quantitatively in ref.[10].

We expect this simple picture to be true for all wavevectors $0 \leq Q \leq
\pi$
since the triplet states with arbitrary $Q$ will be split also by the
anisotropy.
There is thus two distinct magnetic modes throughout the Brillouin zone: one IP
 (in-plane) mode (doubly degenerate) and one OP (out-of-plane) mode.
The IP mode is seen in the $S^z=\pm 1$ sector while the OP mode appears in
the $S^z=0$ sector.
The dispersion of the IP mode has been
obtained for various anisotropies including that of NENP in [23]. INS
experiments
[17-19,22] have resolved these
two modes in the neighborhood of $Q=\pi$: the OP mode is found at 2.5 meV and
the IP mode is even further resolved into two components at 1.05 meV and 1.25
meV. This splitting is due to a small in-plane anisotropy of the type
$E\sum_i [(S_i^x)^2 -(S_i^y)^2]$ that lifts the degeneracy of the doublet
states
$S^z =\pm 1$.
These values of the gaps (ignoring the in-plane anisotropy) leads to
$J=44K$ and $D/J=0.18$ [10]. Detailed interpretation of INS requires a
calculation of the structure factor $S(Q,\omega )$.

\bigskip
\bigskip
\noindent{\bf III. EVALUATING DYNAMICAL QUANTITIES}
\medskip
We describe in this section the methods used to compute the dynamical structure
factor:
$$
S^{\alpha\alpha}(Q,\omega )=\int \! {\rm dt} \quad {\rm e}^{i\omega t}
<0|S^{\alpha}_{-Q} ({\rm t})S^{\alpha}_Q (0)|0>,\quad\quad \alpha ={\rm x,y,z}.
\eqno(3.1)
$$
Here we denote the ground state by $|0>$ and the components of the
Fourier transform of the spin vectors
are $S^{\alpha}_{Q} ({\rm t})$ in the Heisenberg representation.
We first write the structure factor as:
$$
S^{\alpha\alpha}(Q,\omega )=\sum_n
\delta (\omega -(\epsilon_n -\epsilon_0 ))\quad
|\!<n|S^{\alpha}_Q |0>\!|^2 ,
\eqno(3.2)
$$
where the sum over $|n>$ means over all excited eigenstates of the system and
$\epsilon_n$ denotes the energy of $|n>$. In a finite system calculation, one
obtains a finite set of delta functions with weights given by the matrix
elements
appearing in Eq.(3.2). The Lancz\H os
method which is commonly used to obtain the ground state wavefunction and the
first few excited levels is suited to the study of the dynamical properties
[31].

We proceed as follows:

\noindent i) One needs to know first the vector $|0>$. We use the Lancz\H os
algorithm applied to the Hamiltonian H. In fact, any kind of algorithm can
be used at this level.

\noindent ii) One constructs the state:
$$
|\Phi_0 >= S^{\alpha}_Q |0>.
\eqno(3.3)
$$
This state $|\Phi_0 >$ is then used to build a new Lancz\H os sequence of
states
$|\Phi_n >$ by the standard procedure of applying the Hamiltonian to $|\Phi_k
>$
and orthonormalization with respect to the last two vectors $|\Phi_{k} >$
and $|\Phi_{k-1} >$. The Hamiltonian H is then tridiagonal in the basis
$\left\{ |\Phi_n > \right\}$. The nonzero elements of H in this basis are
directly provided by the orthonormalization procedure. We exhaust the
corresponding subspace and store the tridiagonal form of the Hamiltonian.

\noindent iii) One diagonalizes the tridiagonal Hamiltonian by a standard
routine. This leads to a set of eigenenergies $\epsilon_n$ as well as the
eigenvectors whose first coordinate in the Lancz\H os basis are precisely the
overlap matrix elements $<n|\Phi_0>$
that appear in the definition of the structure factor
Eq.(3.2). We thus obtain the weight of each delta
function peak. This has to be done separately for each $Q$ as well as for
$\alpha =x,z$.

In the isotropic D=0 case the ground state is a singlet:
$S^{\alpha}_{Q=0}|0>=0$
and thus $S^{\alpha\alpha}(Q=0,\omega )=0$. The vanishing with $Q$ will occur
quadratically. In the presence of easy-plane
anisotropy the ground state is invariant only
under z rotations: $S^z_{Q=0} |0>=0$ and thus $S^{zz}(Q=0,\omega )=0$,
while $S^{xx}(=S^{yy})$ will be nonzero at $Q=0$ and of order $O(D^2)$. With
in-plane anisotropy $E\sum_i [(S_i^x)^2 -(S_i^y)^2]$
even  $S^{zz}(Q=0)$ will be nonzero and $O(E^2)$.

\bigskip
\bigskip
\vfill
\eject
\noindent{\bf IV. RESULTS FOR $S(Q,\omega )$}
\medskip
We have computed the structure factors $S^{xx}(Q,\omega )$ and
$S^{zz}(Q,\omega )$ for chains of lengths N=4,6,8,10,12,14,16. Some $Q=0$ and
$Q=\pi$ parts of the spectrum have been computed also for N=18.
We have concentrated on the case $D/J=0.18$ quantitatively relevant to
the study of NENP. Our results extend smoothly for not too large
anisotropy $D<J$. For $D=0$ we reproduce previous findings [8] concerning
the lowest excited levels. Previous Monte-Carlo measurements of the
dynamical structure factor [9] for $D=0$ are also compatible with our data.

According to the definition (3.2), we will discuss first the peak positions and
then the corresponding weights. What we observe is that for all fixed
wavevector
$Q$, as a function of $\omega$ the lowest lying peak  concentrates almost all
of the spectral weight. This remark will be made quantitatively precise at
the end of this section. This lowest-lying peak occurs when the frequency
$\omega$ matches the energy of the first excited level with the right quantum
numbers.

In the correlation $S^{xx}$ we find the excitation spectrum from $S^z=0$ to
$S^z=\pm 1$ that we obtained previously [23]. This is plotted in fig.1 as a
function of the momenta by crosses of various sizes. For $Q=0$ and $Q=\pi$
there is enough data to
allow an extrapolation to the thermodynamic limit by use of the Shanks
tranformation. We have also computed the these gaps ($Q=0$ and $Q=\pi$) for
a N=18 chain. The extrapolation leads to $0.301 J$ at $Q=\pi$ and $0.986 J$ at
$Q=0$. In the interval $[0,\pi ]$ there is only the middle of the Brillouin
zone $Q=\pi/2$ where we obtain several data  points from N=4,8,12,16. Shanks
extrapolation leads then to $2.75 J$ for $Q=\pi/2$.

In the correlation $S^{zz}$ we find the excitation spectrum inside the $S^z=0$
subspace. The lowest lying excitation is plotted in fig.1 with empty octogons.
At
$Q=\pi$ the finite-lattice data extrapolate to $0.655 J$. The convergence in
this
case $Q=\pi$ is very good as expected from a massive theory yielding
well-separated eigenvalues in the finite systems [23]. On the contrary at $Q=0$
the convergence is extremely bad and one can only suggest a value of $0.60
\pm 0.05 J$ for this gap. This is clearly due to the presence of a continuum
of states starting immediately above this gap. Such a phenomenon is expected
if one describes the spectrum near $Q=0$ as due to two particles with momenta
near $Q=\pi$ [8].

We note that the excitations in this sector (zz) are above the in-plane
excitations in the neighborhood of $Q=\pi$ but they cross for $Q/\pi\approx
0.75$ and on the top of the dispersion curve it is the in-plane mode which
has highest energy: $2.75 J$ against $2.65 J$ (extrapolated value) for the
out-of-plane mode. This behaviour persists in the region $0\leq Q\leq\pi/2$ of
the Brillouin zone.

To gain understanding of the nature of the excitations we have performed
a variational calculation along the lines proposed by G\'omez-Santos [26].
One discards the states with parallel spins, either if they are
nearest-neighbors or separated by any number of zeros. The typical
states in this subspace are of the form:
$$
|\dots\uparrow\downarrow\uparrow 0\downarrow 00 \uparrow\downarrow\dots >
\eqno(4.1)
$$
There is strict N\'eel ordering of the $S^z=\pm 1$ sites but there can be
any number of intermediate zeros. In this subspace
the true degrees of freedom are then the "spin-zero defects" i.e. sites with
$S^z_i=0$. These domain walls are then represented as fermions (which is a
natural
way of enforcing no double occupancy). The fermionic Hamiltonian can then
be treated by approximate methods. It has been shown that a simple Hartree-Fock
decoupling leads to a good approximation of the spectrum for $D=0$.
The best results are obtained by using the variational improvement
introduced by G\'omez-Santos where one allows a small admixture of states with
parallel spins. The Haldane gap, as computed by this method, is found
to be 0.45J for D=0.
It is straightforward to include anisotropy of the form $D\sum_i (S^z_i)^2$.
We have obtained a spectrum of massive fermions $E(Q)$ that is plotted as a
continuous curve in fig.1. This curve should be the excitation spectrum in the
$S^z=0$ subspace due to the restriction to the subspace (4.1). The gap at
$Q=\pi$
is found to be  $\approx 0.70J$ when D/J=0.18. The single-fermion
curve $E(Q)$ reproduces extremely well the OP dispersion in the region
$\pi/2<Q<\pi$. The IP dispersion will be discussed later on.

\vskip 14.5 truecm
\includegraphics{fig1.ps}

\centerline{\bf Figure 1}

\noindent{\it
Dispersion of magnetic excitations $S^z=0$ and $S^z=\pm 1$ at $D=0.18 J$ for
momenta ranging from 0 to $\pi$. The
continuous line is the result of a variational calculation. The dashed line
is the corresponding edge of two-particle excitations.}

We now turn to the discussion of the spectral weight associated to
the peak positions. For all chain lengths studied for all wavevectors in the
interval $[\approx 0.3-1.0]\times\pi$ we find that both $S^{xx}$ and $S^{zz}$
are
dominated by a single peak as a function of the frequency. This peak
concentrates at least 98 per cent of the spectral weight. Multimagnon
contributions are thus extremely small in the whole range $[\approx
0.3-1.0]\times\pi$. As typical examples we plot in fig.2
$S^{xx}(Q=\pi ,\omega)$+$S^{zz}(Q=\pi ,\omega)$ from the N=16 chain. The peaks
have been broadened for clarity. In fig.3 we plot the $Q=\pi/2$ case showing
the
interchange of the two modes. A single-mode approximation will be an extremely
good description of the physics in this part of the Brillouin zone. The simple
picture described above breaks down for small values of Q. In the N=16 chain,
for
Q=$3\pi/8$ there is still the one-peak structure while for Q=$\pi/4$ there are
several peaks as a  function of $\omega$: see fig.4.
In the N=14 chain we find that the continuum appears already for Q=$2\pi/7$
but is not there for $3\pi/7$.
It is difficult to give a precise value
of the wavevector at which this phenomenon appears since the discretization
of the momenta imposed by the chain length is coarse. As discussed below,
this can be interpreted as evidence for a two-particle continuum.
Our present data suggest that the continuum sets in at $\approx 0.3\pi$ for
both
IP and OP sectors. We see no reason why the continuum boundary should
the same for both modes although we are unable to see any quantitative
difference
in our present data between IP and OP modes in this respect.

Let us discuss the nature of the two-particle excitations. If a single-particle
excitation has a dispersion $E(Q)$, the lower edge of the
continuum of two-particle excitations is given by $E^{(2)}(Q) ={\rm min}_K
(E(K)+E(Q-K))$ in the non-interacting case. Such a free picture seems to apply
to the spin 1 chain [1,2,24].
In the anisotropic system it is possible to construct several continua
since there are two distinct modes (IP and OP).

In the $S^z=\pm 1$ subspace two-particle states are obtained from one
$S^z=\pm 1$ excitation and one $S^z=0$ excitation. The gap of this continuum at
$Q=0$ will be the sum of the IP and the OP gap at $Q=\pi$ i.e.
$\approx 0.301 J + 0.655 J$ in agreement with our extrapolation of 0.986J.
The continuum is also seen in the poor convergence of finite-size data.
Its progressive build-up is seen in fig.4.

In the $S^z=0$ subspace there are two ways of building two-particle states:
either with two $S^z=0$ states or with one excitation $S^z=+1$ and one
excitation $S^z=-1$. In the first case we can obtain an approximation for
the continuum boundary by using the fermionic method cited above:
the curve $E^{(2)}(Q)$ for two fermionic excitations belonging to the $S^z=0$
sector is plotted as a dash-dotted line in fig.1. The
two-fermion continuum is clearly above the OP  (and IP) mode for $0<Q<\pi/2$.
However the second possibility which is the continuum ($S^z=+1$) + ($S^z=-1$)
has
a  Q=0 gap  that is twice the gap at Q=$\pi$ for the IP mode i.e.
$\approx 0.6 J$.  This agrees with our estimate $\approx
0.6 J$ for the gap in the OP sector at Q=0. This continuum is thus the
lowest-lying one. This reasoning shows also that the OP ($S^z=0$)
mode should be below the IP mode for $Q\approx 0$ since the IP gap is then
given
by $\approx 0.301 J + 0.655 J$. In fact, as stressed above, we have
good evidence that the crossing of the two modes occurs quite near
$Q/\pi \approx 0.75$.

\vfill\eject
      ~
\vskip 16.5 truecm
\includegraphics{fig234.ps}
\vskip  1.5 truecm
\centerline{\bf Figure 2,3,4}

\noindent{\it
The structure factor $S^{(xx)}(Q,\omega)  + S^{(zz)}(Q,\omega)$,
for $Q=\pi, \pi/2, \pi/4$ respectively,
as a function of the frequency $\omega$ (in units of J) for chain length N=16.
In fig.2 each of the two functions $S^{(\alpha\alpha)}$ has a single-peak
structure:
a single-mode approximation works very well. The IP mode is doubly degenerate
and appears only in $S^{(xx)}$ while the OP appears in $S^{(zz)}$.In fig.3 the
IP and
OP modes are interchanged with respect to fig.2. In fig.4 a single-mode
approximation no longer holds and the
two-particle continuum is seen: spectral weight is transferred to multimagnon
excitations.}
\vskip 0.5 truecm

The agreement between the {\it ab initio} data and the variational
calculation confirms the relevance of domain walls in the whole Haldane
phase, as is also pointed out in [32]. The disordered-flat phases
obtained in the solid-on-solid picture of the spin-1 chain [33] also
correspond to the subspace of spin-zero defects which is so successfull in
the description of the elementary excitations.

Finally we quote results for the static structure factors
$S^{(\alpha\alpha)}(Q)$
obtained by integration over frequency. In fig.5 these quantities are plotted
in a logarithmic scale as a function of momentum. Due to our limited chain
length, we cannot make any firm statement about the prediction [1] of
square-root Lorentzian behaviour near $Q=\pi$. Our data are compatible with
previous studies [34,35] concluding to a correlation length $\approx 6$ lattice
spacings in the $D=0$ case: the correlation length does not change much when
$D/J=0.18$. The behaviour of S$^{(zz)}$(Q)$\approx Q^2$ sets in
only for very small wavevectors: for the largest part of the Brillouin zone
excepting the neighborhood of $Q=\pi$ it is rather close to $Q^{3/2}$.

\vskip 14.5 truecm
\includegraphics{fig5.ps}

\centerline{\bf Figure 5}

\noindent{\it
The static structure factors $S^{(xx)}$ and $S^{(zz)}$ for all chain lengths
as a function of momentum. The symbol sizes are the same as in figs.1 and 6.
Crosses stand for $S^{(xx)}$ and octogons for $S^{(zz)}$.}
\vfill\eject
\noindent{\bf V. COMPARISON WITH NEUTRON SCATTERING EXPERIMENTS}
\medskip
The two magnetic modes in the neighborhood of $Q=\pi$ have been observed
by INS in ref. [17-20]. The values of the gaps lead to $J\approx 3.8 $meV
and D/J=0.18. These values are very close to those extracted from
magnetization measurements [36]. The dispersion has been studied throughout
the whole Brillouin zone [22]. A good fit is obtained through the
following formula:
$$
\omega_{\pm} (Q)=\sqrt{\Delta^2_{\pm}+v^2\sin^2 (Q)+A^2\cos^2 ({Q\over 2})}.
\eqno(5.1)
$$
The parameters are $\Delta_{+}=2.5$ meV, $\Delta_{-}=1.2$ meV, $v$=9.6 meV and
A=6.1 meV. Scaling by J=4 meV we plot $\omega_{\pm} (Q)$ as continuous lines in
fig.6. For clarity we have also plotted our Lancz\H os points.
The edges of two-particle continuum extracted from the fits (5.1) are also
plotted in fig.6. The lower dashed line corresponds to the ($S^z=+1$) +
($S^z=-1$) continuum while the upper dashed line corresponds to ($S^z=+1$) +
($S^z=0$).
\includegraphics{fig6.ps}

\vskip 12.0 truecm
\centerline{\bf Figure 6}

\noindent{\it
Lancz\H os results compared to experimental fits. The two continuous curves
are taken from Eq.(5.1). They reproduce the findings of ref.22. The two
dash-dotted lines are the two-magnon continuum boundaries obtained also from
(5.1). The size of the symbols is chosen as in fig.1.}

There is very good agreement with the theoretical results
in the neighborhood of Q=$\pi$ where the IP and
OP modes are separated by the experimental resolution [17-20,22].
In the region $\pi/2 <Q<\pi$ the fits (5.1) are also in agreement with
the theoretical results. The two modes are no longer separated since they
are very close to each other. At the top of the dispersion curve Q=$\pi/2$
we estimate from the values quoted in sect. IV the gaps to be 2.75 J for the IP
mode and 2.65 J for the OP. This means
that the IP mode is at $\approx$ 11 meV and the OP mode at $\approx$ 10.5 meV
using the value of J quoted above. The crossing (or near crossing) of the two
modes for $Q\approx 0.75\pi$ precludes their separation in this region of Q.
In the regime $Q<\pi/2$ the separation of the modes increases but the
intensity of the scattering decreases strongly as is seen from fig.5.
The continuum will appear for $Q<0.3\pi$ and is not seen in present
experiments [22]. We see it for Q=$0.25\pi$ and below but there the structure
factor will be very small. For the whole interval $0.3\pi-1.0\pi$ our
data are well reproduced by  long-lived excitations as is seen experimentally
[22]. The asymmetry with respect to $\pi/2$ of the spectrum (figs1,6)
demonstrates the absence of broken translational symmetry. The trend
of the magnetic intensity versus Q in fig.5 is that found by INS [22].
Absence of data for Q very close to $\pi$ forbids us to check the
square-root Lorentzian behaviour expected for each of the static functions
$S^{(xx)}$ and $S^{(zz)}$.

With respect to what has already been done, it would be very interesting
to separate the two modes near $\pi/2$ where intensity is not
dramatically weak. The observation of the two-particle continuum on the
contrary should be quite difficult since it appears only for $Q<0.25\pi$
where magnetic scattering is very weak. The observation of very long-lived
modes [22] is thus in very good agreement with the physics of an anisotropic
spin 1 chain.

Finally we comment on the uncertainties in the numerical data. For the gap
values there is very good convergence for Q=$\pi$ and there one can use
all the chain lengths at our disposal. For other values of Q it is
only at the top of the spectrum that an extrapolation can be performed.
Concerning the values of Q at which something interesting happens (i.e.
crossing
of  the two modes or entering the continuum) without being able to make refined
extrapolations we can only observe that they do not
depend much on the chain length. The corresponding numbers should be regarded
as
tentative (i.e. 0.75$\pi$ for the crossing and 0.3$\pi$ for the continuum).

High resolution INS experiments [30] have revealed that the IP mode is split by
in-plane anisotropy $E\sum_i [(S_i^x)^2 -(S_i^y)^2]$. Our results neglecting
this further splitting thus will apply to the description of actual
experiments as long as the resolution is not very high. This effect
is not expected to change much the numerical figures.

\bigskip
\bigskip
\vfill
\eject
\noindent{\bf VI. CONCLUSION}
\medskip
We have studied the dynamical properties of a spin 1 chain with single-ion
easy-plane anisotropy thought to be relevant to the magnetic behaviour of
the compound NENP. The dynamical structure factor $S^{(\alpha\alpha)}
(Q,\omega )$ has been computed by a Lancz\H os method on chains of lengths
up to 16 spins while some parts of the spectrum have been obtained on a 18
spins chain. We have performed a variational calculation to obtain the
spectrum of elementary excitations in the $S^z=0$ sector. Both {\it ab initio}
and approximate methods agree very well, elucidating the role of the spin-zero
defects in the Haldane phase.

We find that the magnetic excitations are
described by two distinct long-lived modes throughout most of the Brillouin
zone
$0.3-1.0\pi$ where the lowest excited state in each sector ($S^z=0$ and
$S^z=\pm 1$) carries  almost all of the spectral
weight: a single-mode approximation is thus adequate. We find that these modes
merge into a two-particle continuum for $Q<0.3\pi$ as seen in the frequency
dependence of the structure factor. We have shown that these results
reproduce the INS experiments on NENP. The dispersion of the magnetic
modes as well as their intensity is close to that found theoretically.

Our results show that in principle it should be possible to observe the
crossing of the two IP and OP magnetic modes. At the top of the
dispersion curve their separation and intensity should allow observation.
On the contrary the edge of the two-particle continuum appear only
for very small Q$<0.3\pi$ where magnetic intensity is very weak
as found in recent experiments [22]. We mention finally that it would be very
interesting to measure the static form factors of an anisotropic chain
by a method that allows
to reach the neighborhood of Q=$\pi$ since here we are limited by the
coarse-graining due to our small chains. Quantum Monte-Carlo calculation
that works efficiently for static quantities can reach this goal.

\bigskip
\bigskip
\noindent
{\bf ACKNOWLEDGEMENTS}
We thank S. Ma et al. for sending us a copy of ref.22 prior to
publication. Thanks are also due to L. P. Regnault for discussions about
neutron scattering experiements on NENP. We also thank Th. Garel and J. Miller
for reading our manuscript.
\bigskip
\bigskip
\noindent
{\it Note added:} while preparing this manuscript, we received a
preprint by M. Takahashi which addresses similar questions.
\vfill
\eject
\centerline{\bf REFERENCES}
\bigskip
\item{[1]}F. D. M. Haldane, Phys. Rev. Lett. {\bf 50}(1983) 1153; Phys. Lett.
A {\bf 93}(1983) 464.
\medskip
\item{[2]}For a recent review see: I. Affleck, J. Phys. Cond. Matter {\bf 1}
(1989) 3047.
\medskip
\item{[3]}R. Botet and R. Jullien, Phys. Rev. B{\bf 27}, 613 (1983);
M. Kolb, R. Botet and R. Jullien, J. Phys. A{\bf 16}, L673 (1983);
R. Botet, R. Jullien and M. Kolb, Phys. Rev. B{\bf 28}, 3914 (1983),
Phys. Rev. B{\bf 30}, 215 (1984).
\medskip
\item{[4]}J. B. Parkinson and J. C. Bonner, Phys. Rev. B{\bf 32}, 4703 (1985);
J. C. Bonner, J. Appl. Phys. {\bf 61}(8), 3941 (1987).
\medskip
\item{[5]}A. Moreo, Phys. Rev. B{\bf 35}, 8562 (1987).
\medskip
\item{[6]}T. Sakai and M. Takahashi, Phys. Rev. B{\bf 42}, 1090 (1990),
Phys. Rev. B{\bf 42}, 4537 (1990).
\medskip
\item{[7]}M. P. Nightingale and H. W. J. Bl\"ote, Phys. Rev. B{\bf 33}, 659
(1986).
\medskip
\item{[8]}M. Takahashi, Phys. Rev. Lett. {\bf 62}, 2313 (1989).
\medskip
\item{[9]}J. Deisz, M. Jarrell and D. L. Cox, Phys. Rev. B{\bf 42}, 4869
(1990).
\item{[10]}O. Golinelli, Th. Jolic\oe ur and R. Lacaze, Phys. Rev. B{\bf 45},
9798 (1992).
\medskip
\item{[11]}I. Affleck, T. Kennedy, E. H. Lieb and H. Tasaki, Phys. Rev. Lett.
{\bf 59}, 799 (1987); Comm. Math. Phys. {\bf 115}, 477 (1988).
\medskip
\item{[12]}W. J. L. Buyers, R. M. Morra, R. L. Armstrong, P. Gerlach and
K. Hirakawa, Phys. Rev. Lett. {\bf 56}, 371 (1986); R. M. Morra, W. J. L.
Buyers, R. L. Armstrong and K. Hirakawa, Phys. Rev. B{\bf 38}, 543 (1988).
\medskip
\item{[13]}M. Steiner, K. Kakurai, J. K. Kjems, D. Petitgrand and R. Pynn,
J. Appl. Phys. {\bf 61}, 3953 (1987).
\medskip
\item{[14]}Z. Tun, W. J. L. Buyers, R. L. Armstrong, E. D. Hallman and
D. P. Arovas, J. Phys. (Paris) Colloq. {\bf 49}, Suppl. 12, C8-1431 (1988).
\medskip
\item{[15]}Z. Tun, W. J. L. Buyers, R. L. Armstrong, K. Hirakawa and
B. Briat, Phys. Rev. B{\bf 42}, 4677 (1990).
\medskip
\item{[16]}Z. Tun, W. J. L. Buyers, A. Harrison and J. A. Rayne, Phys. Rev.
B{\bf 43}, 13331 (1991).
\medskip
\item{[17]}J. P. Renard, M. Verdaguer, L. P. Regnault, W. A. C. Erkelens,
J. Rossat-Mignod and W. G. Stirling, Europhys. Lett. {\bf 3}, 945 (1987).
\medskip
\item{[18]}J. P. Renard, L. P. Regnault, M. Verdaguer, J. Phys. (Paris),
C8-1425 (1988).
\medskip
\item{[19]}J. P. Renard, M. Verdaguer, L. P. Regnault, W. A. C. Erkelens,
J. Rossat-Mignod, J. Ribas, W. G. Stirling and C. Vettier, J. Appl. Phys.
{\bf 63}(8), 3538 (1988).
\medskip
\item{[20]}K. Katsumata, H. Hori, T. Takeuchi, M. Date, A. Yamagishi
and J. P. Renard, Phys. Rev. Lett. {\bf 63}, 86 (1989).
\medskip
\item{[21]}A. Meyer, A. Gleizes, J.-J. Girerd, M. Verdaguer and O. Kahn,
Inorg. Chem. {\bf 21}, 1729 (1982).
\medskip
\item{[22]}S. Ma, C. Broholm, D. H. Reich, B. J. Sternlieb and R. W. Erwin,
"Dominance of long-lived excitations in the antiferromagnetic spin-1 chain
NENP", preprint 1992.
\medskip
\item{[23]}O. Golinelli, Th. Jolic\oe ur and R. Lacaze, Phys. Rev. B{\bf 46},
 10854 (1992).
\medskip
\item{[24]}A. M. Tsvelik, Phys. Rev. B{\bf 42}, 10499 (1990).
\medskip
\item{[25]}I. Affleck and R. A. Weston, Phys. Rev. B{\bf 45}, 4667 (1992).
\medskip
\item{[26]}G. G\'omez-Santos, Phys. Rev. Lett. {\bf 63}, 790 (1989).
\medskip
\item{[27]}E. H. Lieb and D. C. Mattis, J. Math. Phys. {\bf 3}, 749 (1961);
D. C. Mattis, ``The Theory of Magnetism'', Vol. I and II, Springer
Verlag (Berlin) 1985.
\medskip
\item{[28]}C. M. Bender and S. A. Orszag, ``Advanced Numerical Methods for
Scientists and Engineers'', MacGraw and Hill (New York) 1987.
\medskip
\item{[29]}F. R. Gantmacher, ``Matrix Theory'', Chelsea (New York) 1964.
\medskip
\item{[30]}L. P. Regnault, J. Rossat-Mignod and J. P. Renard, J. Mag. Mag.
Mater, {\bf 104}, 869 (1992).
\medskip
\item{[31]}E. Gagliano and C. Balseiro, Phys. Rev. Lett. {\bf 59}, 2999 (1987).
\medskip
\item{[32]}H. J. Mikeska, Europhys. Lett. {\bf 19}, 39 (1992); N. Elstner
and H. J. Mikeska, preprint 1992.
\medskip
\item{[33]}M. den Nijs and K. Rommelse, Phys. Rev. B{\bf 40}, 4709 (1989).
\medskip
\item{[34]}M. Takahashi, Phys. Rev. B{\bf 38}, 5188 (1988).
\medskip
\item{[35]}K. Nomura, Phys. Rev. B{\bf 40}, 2421 (1989);
S. Liang, Phys. Rev. Lett. {\bf 64}, 1597 (1990).
\medskip
\item{[36]}T. Delica, K. Kopinga, H. Leschke and K. K. Mon, Europhys. Lett.
{\bf 15}, 55 (1991).
\bye